\documentclass[%
 reprint,
 amsmath,amssymb,
 aps,
pra,
floatfix,
]{revtex4-2}

\usepackage{xcolor}
\usepackage{graphicx}
\usepackage{dcolumn}
\usepackage{bm}
\usepackage{multirow}
\usepackage{booktabs}

\begin{document}
\preprint{APS/123-QED}

\title{Topologically Protected Surface Acoustic Wave Edge States in Phononic Integrated Circuits}

\author{Jack Guida}
\author{Niccol\'o Scalise Pantuso}%
\author{Cristian Cassella}%
\author{Sunil Mittal}%
\author{Siddhartha Ghosh}%
    \email{s.ghosh@northeastern.edu}

\affiliation{Northeastern University, Boston, Massachusetts, USA
}%

\date{\today}
\begin{abstract}
Phononic integrated circuits enable the development of compact systems to manipulate radio frequency signals through interconnected networks of acoustic waveguides. Introducing topologically protected robustness in these circuits to counter fabrication imperfections and local perturbations, remains a challenge. This work presents a validated modeling framework for topologically protected edge states using surface acoustic waves in etch-defined waveguides consisting of scandium aluminum nitride on silicon carbide, directly comparing analytical, full-wave numerical simulation, and laser Doppler vibrometer measured experimental approaches and their agreement. This work represents the first experimental demonstration of topological edge states on a ScAlN on SiC platform, validated through direct quantitative comparison with analytical and numerical analyses, with future implications for topological protection in phononic and optomechanical integrated circuit platforms.
\end{abstract}

\maketitle


\section{\label{sec:Intro} Introduction}
Compact classical signal processing systems often depend on networks of interconnected components. Piezoelectric transduction of electromagnetic waves into the mechanical domain enables these networks to be built on a single solidly mounted platform, substantially reducing device footprint and becoming a foundation of RF signal processing \cite{hashimoto_surface_2000}. Surface acoustic wave (SAW) systems built on such platforms support radio frequency (RF) filters \cite{yang_sv-saw_2025}, full-duplex systems \cite{guida_microwave_2026}, and biomedical/chemical sensing \cite{ji_aptamer-based_2020,xiong_high_2021}. Extending these devices into interconnected, multi-element waveguide architectures has led to the development of phononic integrated circuits (PnICs) and construction of compact front-end systems from connected acoustic elements rather than single and discrete components \cite{guida_phononic_2025,hackett_towards_2021}. 
Realizing these platforms requires an acoustic impedance mismatch between the piezoelectric film and its substrate to yield SAW modes that mitigate leakage into the substrate. Scandium aluminum nitride (ScAlN) films have gained popularity for such platforms due to their high piezoelectric coefficients, in-plane isotropic crystal nature, and CMOS compatibility \cite{caro_piezoelectric_2015}. The high acoustic impedance contrast between ScAlN and silicon carbide has established ScAlN-on-silicon carbide (SoSiC) as a promising platform for on-chip RF signal processing \cite{guida_focused_2024}.

As PnIC architectures grow in size and complexity to support more intricate signal processing functions, robustness against local defects becomes increasingly important. The development of the focusing interdigitated transducer (FIDT) has enabled a Gaussian-beam wavelength-scale focusing of excited SAW modes. The subsequent injection of this focused mode into etch-defined acoustic ridge waveguides enables full in-plane routing and control, with components for signal routing and distribution.
Exploiting topology offers a clear route to such robustness. First studied in the quantum regime \cite{hasan_colloquium_2010}, topological physics over the past decade has attracted growing attention in both acoustics and optics \cite{xue_topological_2022,lu_topological_2014}, enabling extreme wave manipulation and exotic phenomena such as edge and corner states \cite{yang_topological_2015,cha_experimental_2018,xue_acoustic_2019}, backscattering-immune waveguides \cite{wang_observation_2009,wang_extended_2022}, and enhanced phonon-photon interactions \cite{ren_topological_2022,matheny_enhanced_2018}. Among these approaches, the one-dimensional Su-Schrieffer-Heeger (SSH) model provides a simple yet effective means of realizing topological protection in lattice systems \cite{su_solitons_1979}. Topological lattices support edge states with strong localization whose presence is established by the bulk geometry itself and has been demonstrated \cite{de_ponti_localized_2024} in photonic \cite{kraus_topological_2012} and acoustic domains \cite{xu_gigahertz_2025}, with applications for enhanced sensing \cite{maggioli_topologically_2025,Wen_TopoProtected,Kaya_Inertial}. These edge states are inherently protected against local perturbations \cite{liu_rainbow_2023,Wang}, making them extremely robust. Despite the promise shown in other domains \cite{Alu2025TopologicalAcoustics,Puchnin_ApplicationTopo}, topological concepts remain largely unexplored in the context of PnICs. This gap becomes increasingly important as SAWs move beyond classical RF signal processing toward quantum \cite{manenti_circuit_2017} and coherent phononic applications, including recent demonstrations of coherent phonon generation through an electrically injected SAW phonon laser \cite{wendt_electrically_2026}. As SAW platforms are developed toward quantum and coherent applications,  disorder immunity becomes more important in device design.

\begin{figure*}[t!]
\centering
\includegraphics[width=\linewidth]{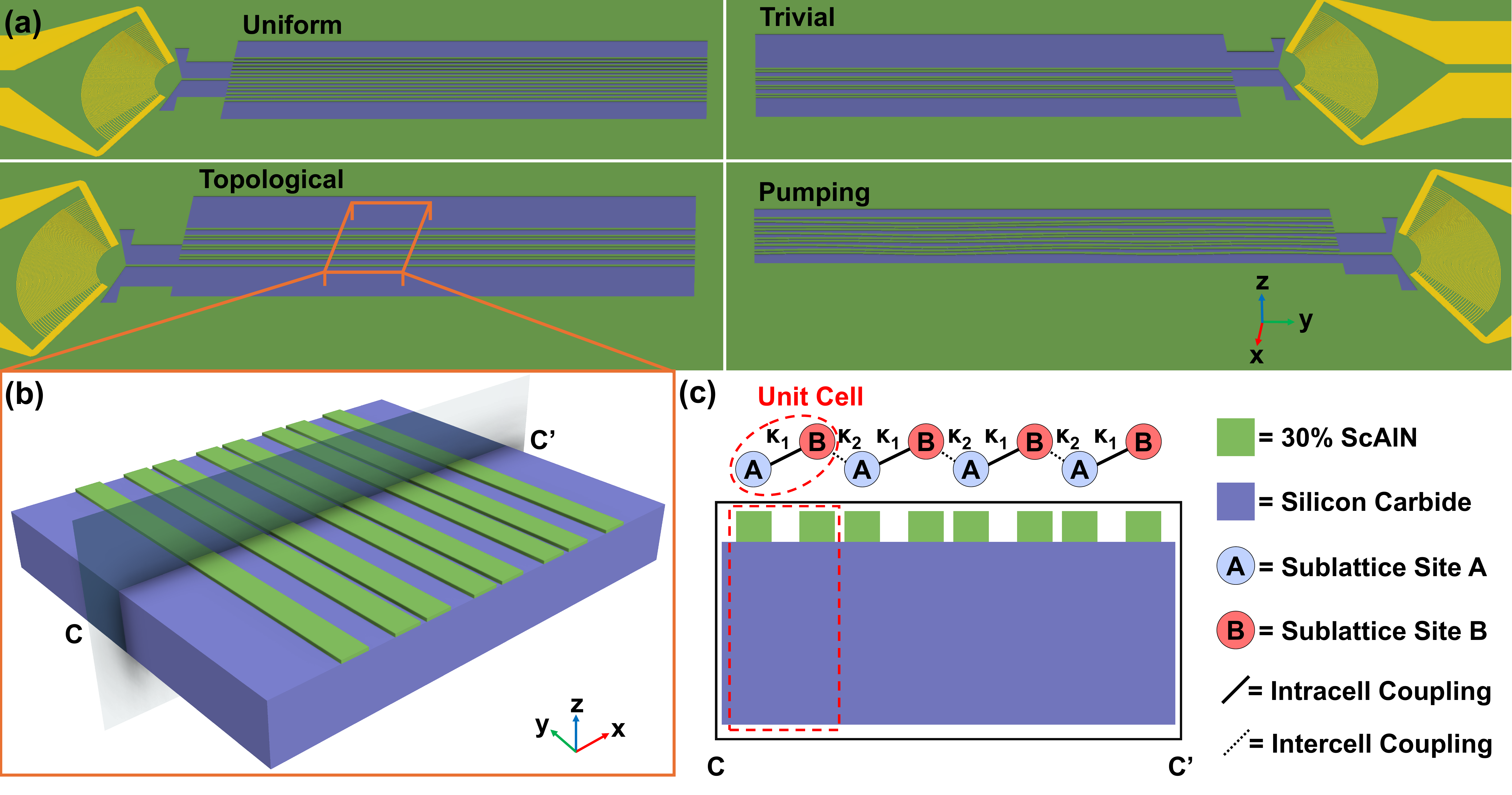}
\caption{(a) Overview of the four phononic coupled waveguide lattice configurations, including uniform, trivial, topological, and Thouless pumping structures. (b) Zoomed view of the topological (SSH) configuration, consisting of eight coupled ridge waveguides with alternating weak/strong coupling gaps, implemented as an acoustic analog of a propagating SAW SSH model with $\kappa_1<\kappa_2$.  (c) Cross-section along cut plane CC’, overlaid with the corresponding toy-model lattice, highlighting intracell vs intercell couplings.}
\label{fig:Overview}
\end{figure*}

Merging topological protection with PnIC architectures offers a path toward waveguide networks with robustness against disorder. This work realizes topological protection in etch-defined piezoelectric SAW waveguide in SoSiC, validated across analytical, numerical, and experimental approaches. A numerical model of coupled acoustic waveguides is developed to calibrate the analytical coupled mode framework, which is then used to design a full topological lattice and its trivial counterpart. Both geometries are implemented and verified through full-wave finite element method simulation in COMSOL. All three methods are then directly compared, with experimental results confirming strong localization of the topologically protected edge states. This work establishes the design and characterization foundation for topologically protected PnICs for robust acoustic signal processing in next generation RF systems.

\section{\label{sec:Concept} Coupled Acoustic Waveguide Analysis}

\begin{figure}[b]
\centering
\includegraphics[width=\linewidth]{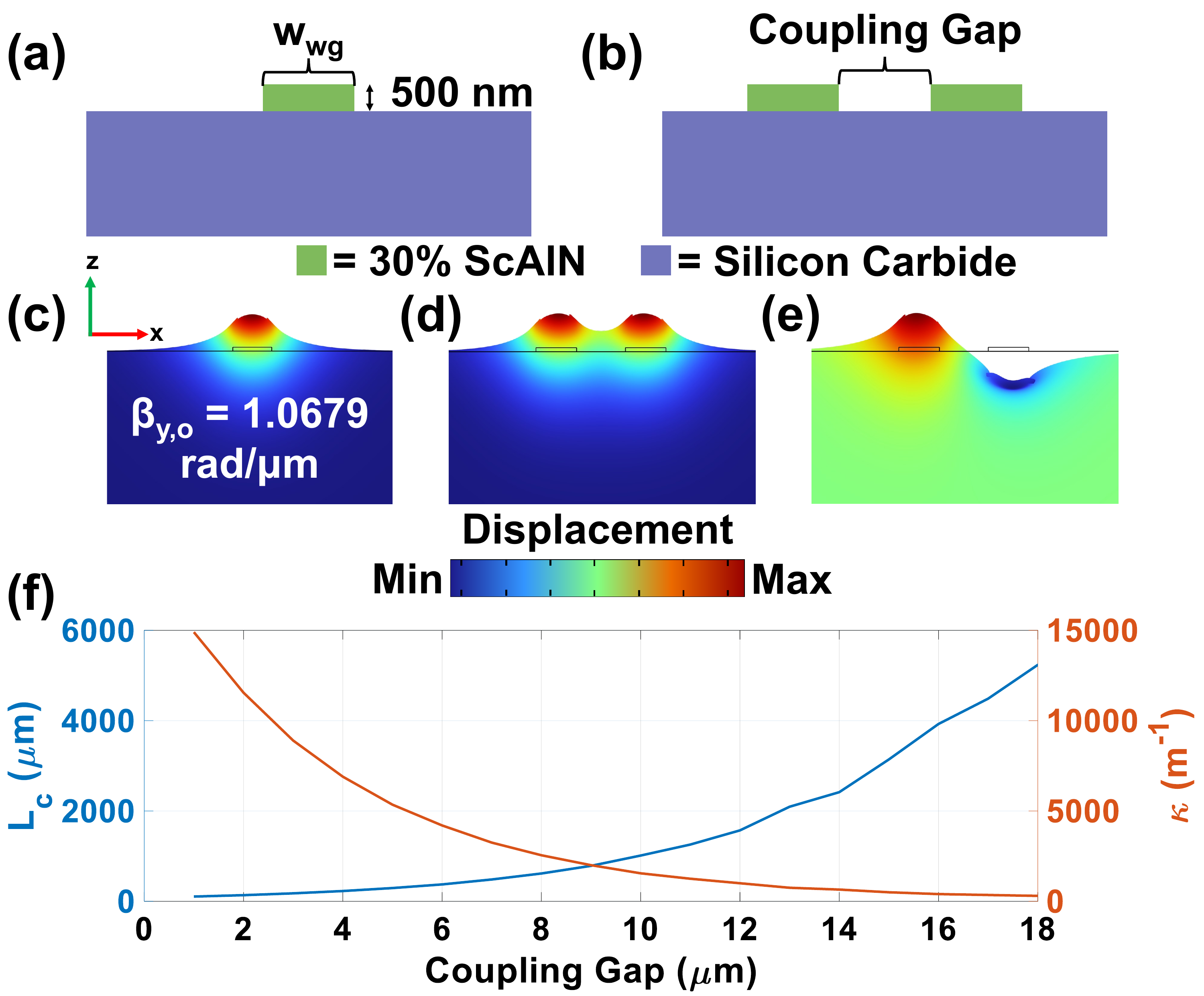}
\caption{(a) Single ridge acoustic waveguide geometry in 500 nm 30\% ScAlN on silicon carbide, with the waveguide width ($w_{wg}$) equal to the acoustic wavelength of interest ($\lambda$ = 5 $\mu$m). (b) Coupled waveguide geometry consisting of two identical waveguides from (a), separated by a coupling gap, g. COMSOL mode analysis of the (c) single ridge waveguide and its wavenumber solution ($\beta_{y,o}$), and (d) even and (e) odd modes of the coupled waveguides, each showing the z-displacement field. (f) Coupling length ($L_c$) and coupling coefficient ($\kappa$) plotted vs. coupling gap, extracted from the even and odd mode wavenumber.}
\label{fig:CouplingAnalysis}
\end{figure}
Introduced in Fig.~\ref{fig:Overview}(a), this work examines four coupled waveguide lattice configurations for PnICs, including uniform, trivial, topological, and Thouless pumping structures. Each configuration is realized using an identical FIDT injection into a system of coupled ridge waveguides, with the specific coupling gap pattern between waveguides (and the choice of injected waveguide) distinguishing the different cases. Fig.~\ref{fig:Overview}(b) shows a magnified view of the topological configuration, consisting of a system of $N$ coupled waveguides where $\kappa_1 < \kappa_2$ and $N$=8 in the ScAlN on SiC platform. The SSH model here serves as the underlying Hamiltonian framework for all four configurations. To understand the coupling dynamics, the cross sectional area is shown and overlaid with the toy-model in Fig.~\ref{fig:Overview}(c), highlighting the alternating intracell and intercell coupling ($\kappa_1$ and $\kappa_2$, respectively) between sublattice sites A and B. 

Coupled waveguides in SAW platforms have been shown to effectively control energy transfer within phononic circuits. Numerical methods can be utilized to extract the relationship of the coupling length ($L_c$) and coupling coefficient ($\kappa$) over the gap ($g$) between the coupled waveguides. The material stack in this work is defined as 500 nm 30\% ScAlN on SiC, with an operating acoustic wavelength ($\lambda_A$) of 5 $\mu$m. Acoustic waveguides often exist in the form of rib, strip loaded, slot loaded, and ridge waveguides (focus of this work) \cite{feng_gigahertz_2023}, where a full etch of the ScAlN defines the acoustic waveguides. Here, the waveguide width is set equal to the acoustic wavelength, ensuring higher-order lateral modes are cutoff. Further details on the numerical simulation can be found in Section I of the Supplementary Material. The geometry for the ridge waveguide can be seen in Fig.~\ref{fig:CouplingAnalysis}(a). Similarly, the coupled waveguide geometry is defined with symmetric waveguides separated by a coupling gap ($g$), as shown in Fig.~\ref{fig:CouplingAnalysis}(b).

To reduce the computational complexity of the numerical modeling, the transducer response is not simulated, but rather a boundary mode analysis is run on the input face of the structure to mimic the injection of the focused SAW mode generated from the FIDT. The fundamental Rayleigh mode (dominated with its vertical displacement) solution in the single-ridge waveguide is shown in Fig.~\ref{fig:CouplingAnalysis}(c), along with its out-of-plane wavenumber solution, $\beta_{y,o}$. In the case of a coupled waveguide system, both even (symmetric) and odd (anti-symmetric) mode solutions exist, as shown in Fig.~\ref{fig:CouplingAnalysis}(d) and (e), respectively. In the total field of the directional coupler, there is a linear combination of field patterns that have different propagation constants, $\beta_{even}$ and $\beta_{odd}$, known as the supermodes of the structure. The maximum power transfer between waveguides occurs when the two modes have a $\pi$ phase shift between them, leading to a coupling length of $L_c = \frac{\pi}{\beta_{even}-\beta_{odd}}$. The coupling coefficient ($\kappa$) can subsequently be extracted from $\kappa = \frac{\pi}{2L_c}$ \cite{chen_study_2026,haus_coupled-mode_1987}. The gap between the waveguides is swept in the numerical simulation and the trend of coupling length and coefficient is visualized over the coupling gap in Fig.~\ref{fig:CouplingAnalysis}(f).

Coupled mode theory provides a fast and intuitive baseline for designing coupled-waveguide architectures, predicting coupling from design parameters \cite{huang_coupled-mode_1994}. This method relies on assumptions that significantly simplify the underlying wave dynamics, and its predictions alone are insufficient to capture the full physical device response. Accurately modeling the full physical structure prior to fabrication is therefore crucial to bridge the gap between analytical design and real device performance. This is especially important for SAW-based lattice designs, where coupled mode theory makes assumptions that do not fully capture the underlying propagation dynamics of the acoustic waveguides, such as diffraction and heterostructure induced losses \cite{slobodnik_surface_1976}, that become significant at wavelength-scale waveguides.

Using coupled mode theory \cite{yariv_photonics_2007}, the two-waveguide system propagating along \textit{y} can be mathematically described with Eq.~\ref{eq:twoWaveguide}, where $a_{n}$ is the amplitude in waveguide $n$.

\begin{equation}
    \label{eq:twoWaveguide}
    \begin{split}
        j\frac{da_1}{dy}&=\beta_{y,o} a_1 + \kappa a_2, \\
        j\frac{da_2}{dy}&=\beta_{y,o} a_2 + \kappa a_1
    \end{split}
\end{equation}

This can be generalized to a system of N-coupled waveguides, where for any waveguide ($n=1:N-1$), its amplitude changes due to its own propagation and the coupling between its nearest neighboring waveguides. This is described by Eq.~\ref{eq:NWaveguide}.

\begin{equation}
    \label{eq:NWaveguide}
    j\frac{da_n}{dy}=\beta_{y,o} a_n + \kappa_{n-1,n} a_{n-1} + \kappa_{n,n+1}a_{n+1}
\end{equation}

Here, an alternating nature in the coupling between waveguides can be introduced, indicated with $\kappa_1$ and $\kappa_2$. Collecting the mode amplitude of each waveguide into a vector, $\vec{a}$, allows the system to be written in terms of a system Hamiltonian, $\mathbf{{H}}$, shown in Eq.~\ref{eq:Hamiltonian}.

\begin{equation}
    \label{eq:Hamiltonian}
    \begin{split}
        j\frac{d\vec{a}}{dy} &= \left(\beta_{y,o} \mathbf{I} + \mathbf{H}\right)\vec{a}, \\
        \mathbf{H} &= \begin{pmatrix}
            0        & \kappa_1 & 0        & 0        & \cdots \\
            \kappa_1 & 0        & \kappa_2 & 0        & \cdots \\
            0        & \kappa_2 & 0        & \kappa_1 & \cdots \\
            0        & 0        & \kappa_1 & 0        & \cdots \\
            \vdots   &          &          &          & \ddots
        \end{pmatrix}
    \end{split}
\end{equation}

Labeling each pair of adjacent waveguides with sub-lattice sites \textit{A} and \textit{B} the Hamiltonian can equivalently be expressed in bra-ket notation, shown in Eq.~\ref{eq:H_braket}, where $\kappa_1$ and $\kappa_2$ represent the intracell and intercell coupling, respectively. 

\begin{equation}
\label{eq:H_braket}
    \begin{split}
        \mathbf{{H}} &= \kappa_1 \sum_{m=1}^{N}{\left( |m,B\rangle \langle m,A| + \text{h.c.}\right)} \\
        &+ \kappa_2 \sum_{m=1}^{N-1}{\left( |m+1,A\rangle \langle m,B| + \text{h.c.}\right)}
    \end{split}
\end{equation}

The Hamiltonian formalism developed and the extracted coupling dynamics provide the analytical foundation for each of the four lattice configurations introduced. This framework is used to derive closed-form analytical solutions for the uniform, trivial, topological, and Thouless pumping structures, enabling direct comparison against numerical and experimental approaches.

\section{\label{sec:Fabrication and Test Setup} Device Fabrication and Transducer Characterization}

\begin{figure}[b]
\centering
\includegraphics[width=\linewidth]{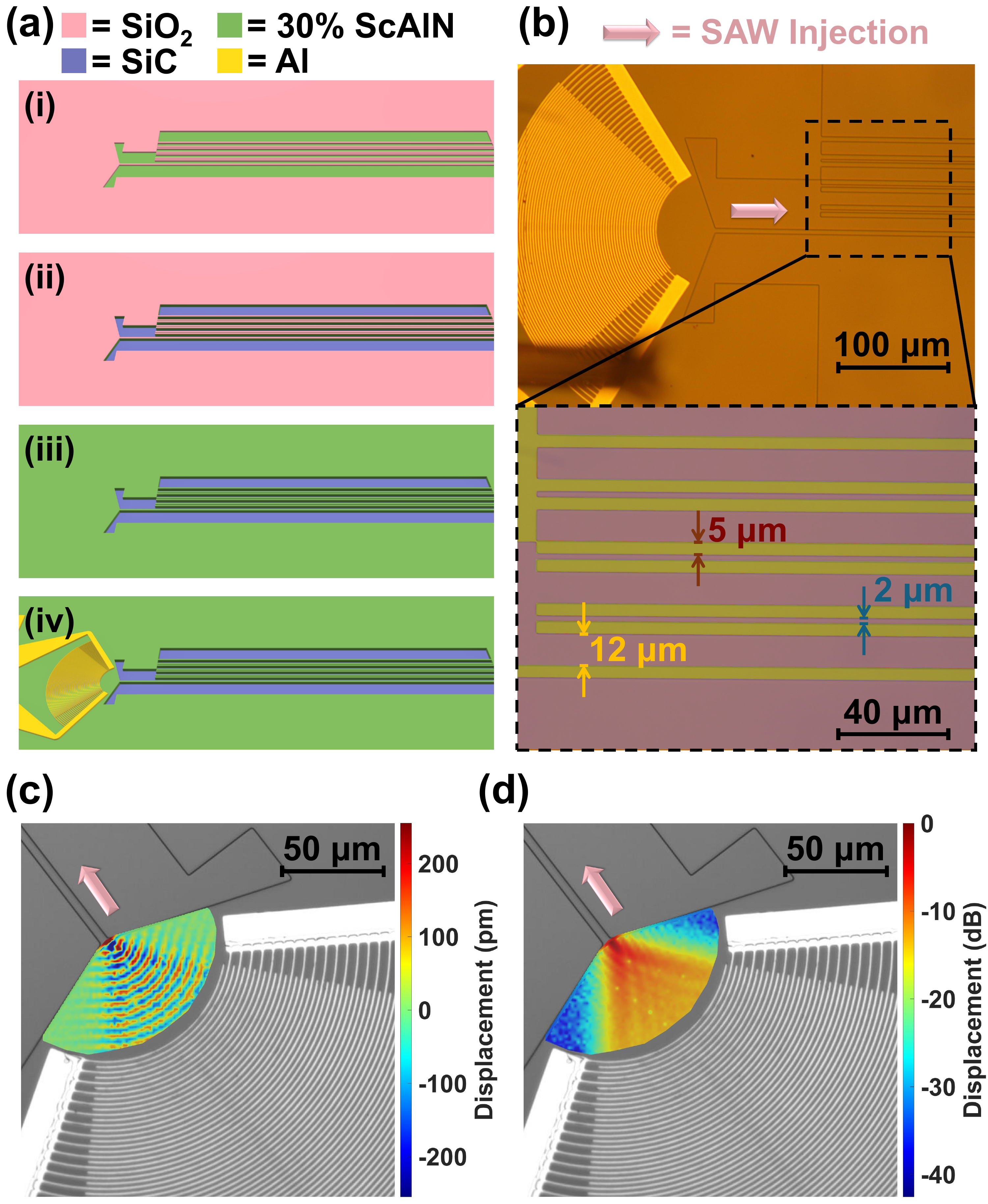}
\caption{(a) Fabrication process of the coupled acoustic waveguide systems, including uniform-coupling, topological, and trivial configurations. (b) Optical image with the inset showing a false-colored image of the topological lattice. LDV-measured characterization of transducer's focusing response, showing (c) real out-of-plane displacement and (d) normalized displacement magnitude (dB).}
\label{fig:Fab}
\end{figure}

Fabrication of the phononic devices is shown in Fig.~\ref{fig:Fab}(a), and starts with the deposition of 500 nm 30\% ScAlN on a SiC substrate using RF magnetron sputtering. ScAlN is an extremely chemically resistant and hard material \cite{shifat_high-accuracy_2024}, and thus a 500 nm layer of silicon dioxide (SiO$_2$) is deposited using plasma-enhanced chemical vapor deposition (PECVD) to act as a hard mask for the etch of the ScAlN. The oxide is patterned using optical lithography and etched utilizing an Ar/CF$_4$/CHF$_3$/H$_2$-based reactive ion etch (RIE). Next, the acoustic waveguides are patterned in the ScAlN using a Cl$_2$/BCl$_3$/N$_2$-based RIE. The oxide hard mask is then removed using a buffered oxide etch (BOE). Finally, the transducers are defined using optical lithography before 100 nm of aluminum is sputtered and lifted-off. An optical image of a fabricated device with a topological configuration is shown in Fig.~\ref{fig:Fab}(b), with the inset highlighting the nature of the coupled waveguides through a false-colored image.

The performance of the FIDTs is crucial to the overall system, as it is responsible for the injection of the focused SAW mode into the waveguide system. Each system uses an identical transducer to inject the mode. To this extent, the FIDT response is first individually characterized to ensure sufficient injection of the fundamental Rayleigh mode. Since SAWs are dominated by their out-of-plane displacement, a laser Doppler vibrometer (LDV) serves as an ideal tool to image the devices and capture the displacement field to characterize the acoustic focusing dynamics. Devices are mounted on a custom PCB and wirebonded before being mounted in the LDV measurement system. A single +21 dBm tone of 1.292 GHz (corresponding to the resonance frequency of the fundamental Rayleigh mode) is input to the transducers, operating with an acoustic wavelength of 5 $\mu$m. The focusing region of the transducer is rastered with a mesh size corresponding to $\lambda_A/4$ and the real component and magnitude of the measured out-of-plane displacement is shown in Fig.~\ref{fig:Fab}(c) and (d), respectively. 

These results validate the LDV as a viable method for measuring the displacement and mode presence throughout the system, while confirming the high-intensity acoustic Gaussian beam focusing and mode injection into an acoustic waveguide. Following the validation of the FIDT and injection of the fundamental Rayleigh mode into the acoustic waveguide, the LDV is used to measure the trivial and topological lattice structures, as well as the Thouless pumping structures to experimentally verify the theoretical and numerical predictions of mode localization and transport, with the experimental configuration described in Section II of the Supplementary Material.

\section{Coupled Waveguide Lattice Configurations}
The analytical approach is extended to four different coupling configurations including uniform coupling and spreading, trivial dimerization, topological dimerization (SSH model), and Thouless pumping. Within each configuration, analytical predictions are directly compared against full wave numerical simulation and experimental measurement, isolating the role of coupling symmetry and modulation in governing edge state localization and transport.

\subsection{Uniform Coupling Lattice}
To establish a baseline, the uniform coupling case ($\kappa_1=\kappa_2$) is first considered, where there is no dimerization and the injected power spreads into bulk modes throughout the coupled waveguide system. Using the information from Fig.~\ref{fig:CouplingAnalysis}(f), an arbitrary gap of 3 $\mu$m was used. To realize the symmetric spreading of the injected mode into the bulk of the system, a lattice of 13 coupled waveguides are analyzed analytically. Eq.~\ref{eq:Hamiltonian} is a linear, constant coefficient ordinary differential equation that can be solved with a closed-form solution, indicated by Eq.~\ref{eq:analyticalSolution}. Since the Hamiltonian is Hermitian (total energy equals its own transpose conjugate), it can be diagonalized as $\mathbf{H = VDV^{-1}}$ where $\mathbf{V}$ is the matrix of eigenvectors and $\mathbf{D}$ is the diagonal matrix of eigenvalues.

\begin{equation}
\label{eq:analyticalSolution}
    \begin{split}
    \vec{a}\left(y\right) &= e^{-j\beta_y y}e^{-j\mathbf{H}y}\vec{a}\left(0\right), \\
    e^{-j\mathbf{H}y}&=\mathbf{V}e^{-j\mathbf{D}y}\mathbf{V^{-1}}
    \end{split}
\end{equation}

\begin{figure}[b]
\centering
\includegraphics[width=\linewidth]{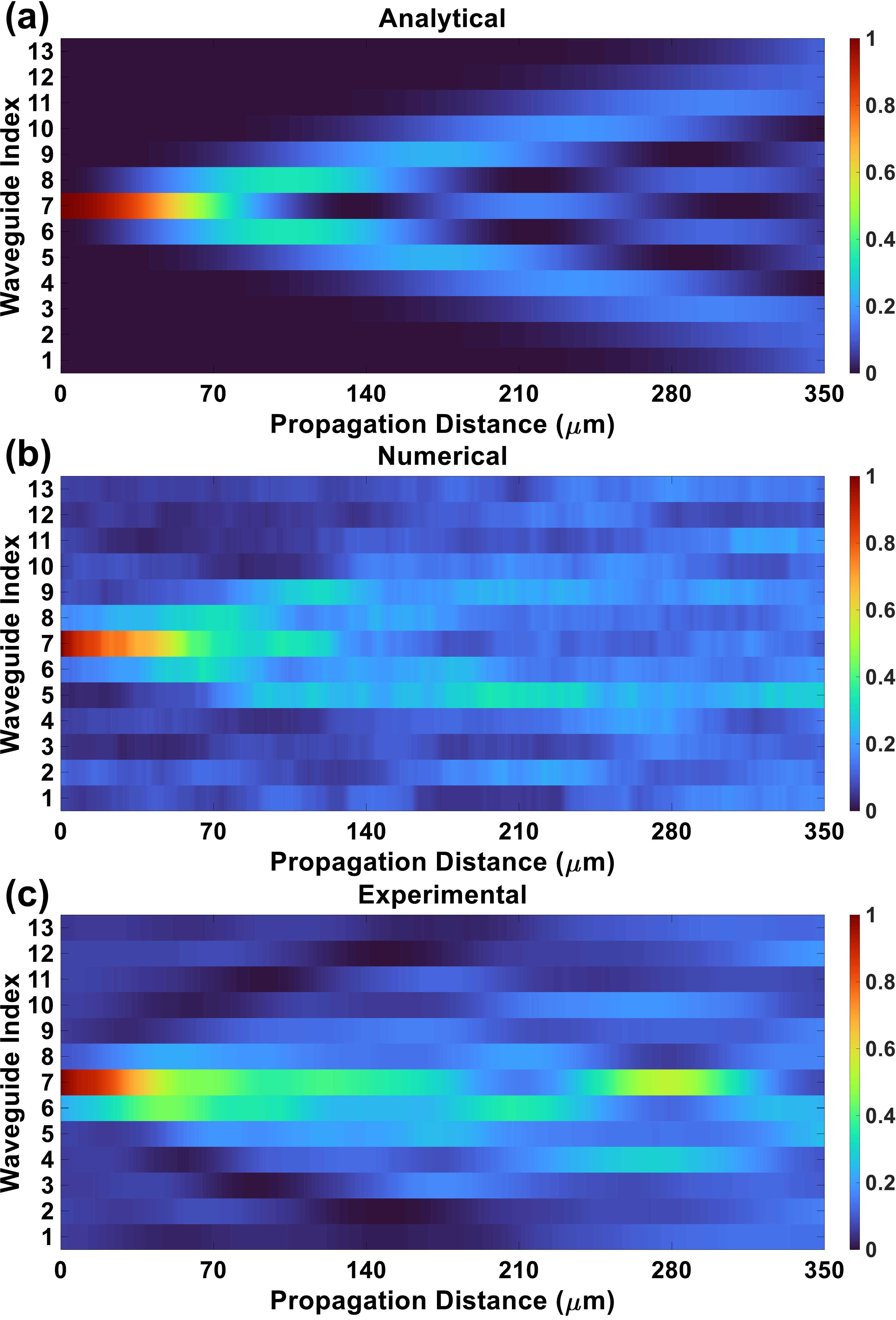}
\caption{Spatial intensity evolution map in the uniform coupling ($\kappa_1 = \kappa_2$) lattice, consisting of 13 coupled waveguides with excitation injected at the center waveguide (Waveguide Index = 7), shown for (a) analytical, (b) numerical, and (c) experimental results.}
\label{fig:spreadingIntensities}
\end{figure}

The analytical solution for a center-waveguide injection is shown in Fig.~\ref{fig:spreadingIntensities}(a). Here, and everywhere to follow, phase oscillation has unit amplitude and is ignored to evaluate the mode amplitude. This beam spatial intensity evolution map illustrates the mode amplitude in each waveguide over a 350 $\mu$m propagation length and the results spreading of power into the bulk. For direct comparison, the system is additionally modeled in COMSOL using a full 3D numerical solution, where the excitation is introduced through the injection of the fundamental Rayleigh mode into the center waveguide from a boundary mode analysis at the input and output, with wave propagation solved throughout the geometry domain. Within the solution, a cut line is defined along the direction of propagation in the center of each waveguide in the system. For each cut line, the mode amplitude is extracted and visualized in Fig.~\ref{fig:spreadingIntensities}(b). While the analytical and numerical solutions show agreement in the trend of spreading, minor deviations occur, likely from dispersive effects and minimal amounts of substrate leakage not captured by the idealized coupled mode model. However, the general spreading into the bulk of the system is captured. 

For each fabricated device, a scan region is defined over the area occupied by the waveguides, and a central cut line along the direction of propagation is used to extract the experimental spatial intensity evolution, consistent with the cut-line approach used in the analytical and numerical analyses. Experimentally, beginning with the uniform spreading configuration, the transducer is driven with the same power and tone as done with the FIDT characterization. The LDV is used to measure out-of-plane displacement where a 20x objective lens is used in the configuration, allowing to capture a propagation distance of roughly 500 $\mu$m in a single scan. In this configuration with device parameters exactly matching that of the theoretical and numerical approaches, the central waveguide in a 13-coupled waveguide system is injected. It is clear that the injected mode starts with the highest intensity, and quickly begins coupling to neighboring waveguides, ultimately spreading into the bulk of the system. The normalized mode intensity over this propagation length is plotted in Fig.~\ref{fig:spreadingIntensities}(c) for a direct comparison with the analytical and numerical predictions. The trends in the case of uniform spreading into the bulk show good agreement with that of the numerical simulation, giving a preliminary indication of accurate modeling of the coupled waveguide systems. 

\subsection{Trivial Lattice}
\begin{figure}[t]
\centering
\includegraphics[width=\linewidth]{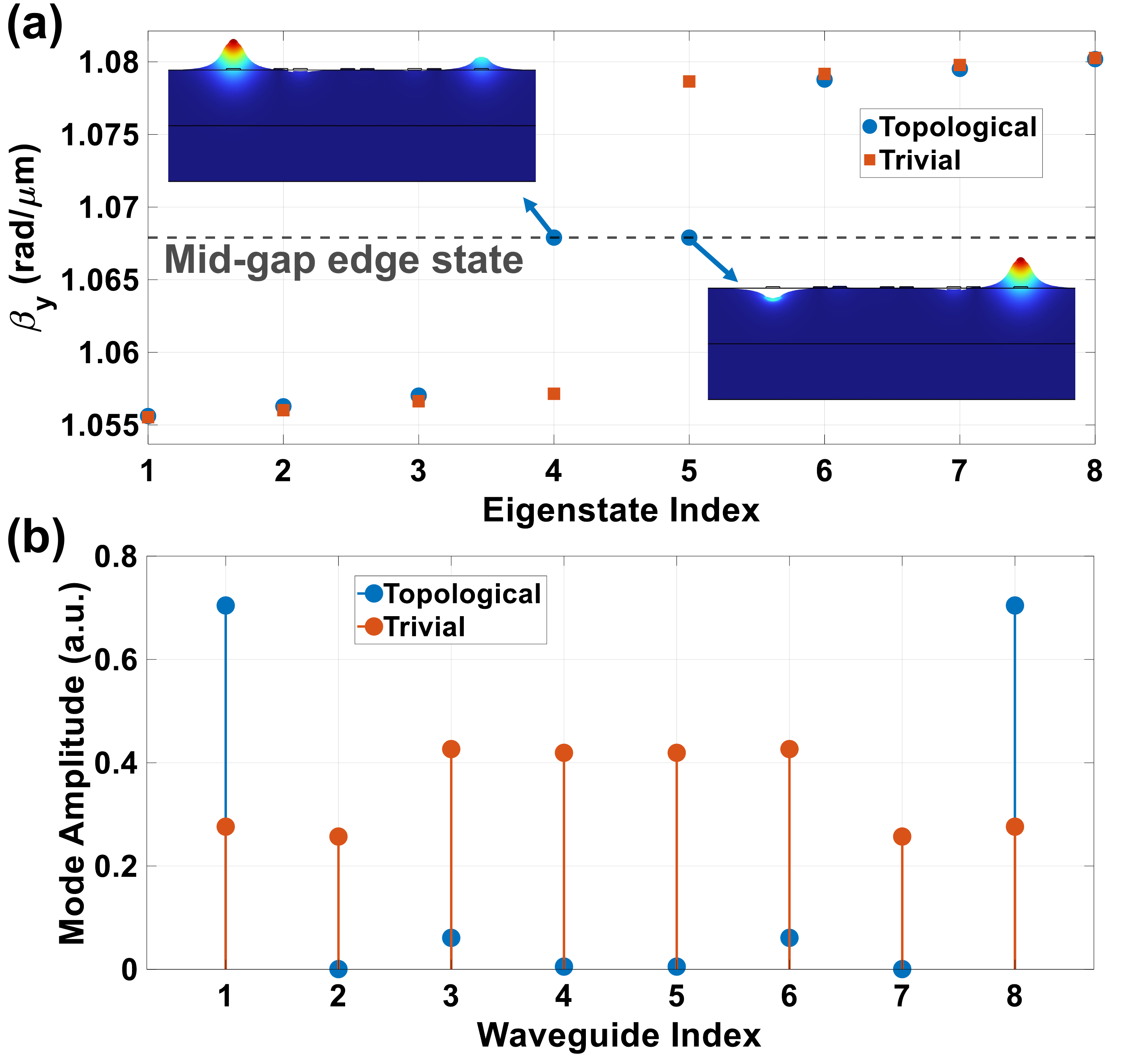}
\caption{(a) Propagation constant spectrum comparing the topological ($\kappa_1 < \kappa_2$) and trivial ($\kappa_1 > \kappa_2$) configurations, with the eigenvalues of the Hamiltonian converted to physical wavenumbers, $\beta_y$. Its corresponding COMSOL-simulated mode profile for the edge states are shown as the inset. (b) Mode amplitude is plotted for each waveguide index in both the topological and trivial cases, showing strong mode localization to the edge waveguides in the topological case.}
\label{fig:eigenstates}
\end{figure}
\begin{figure}[b]
\centering
\includegraphics[width=\linewidth]{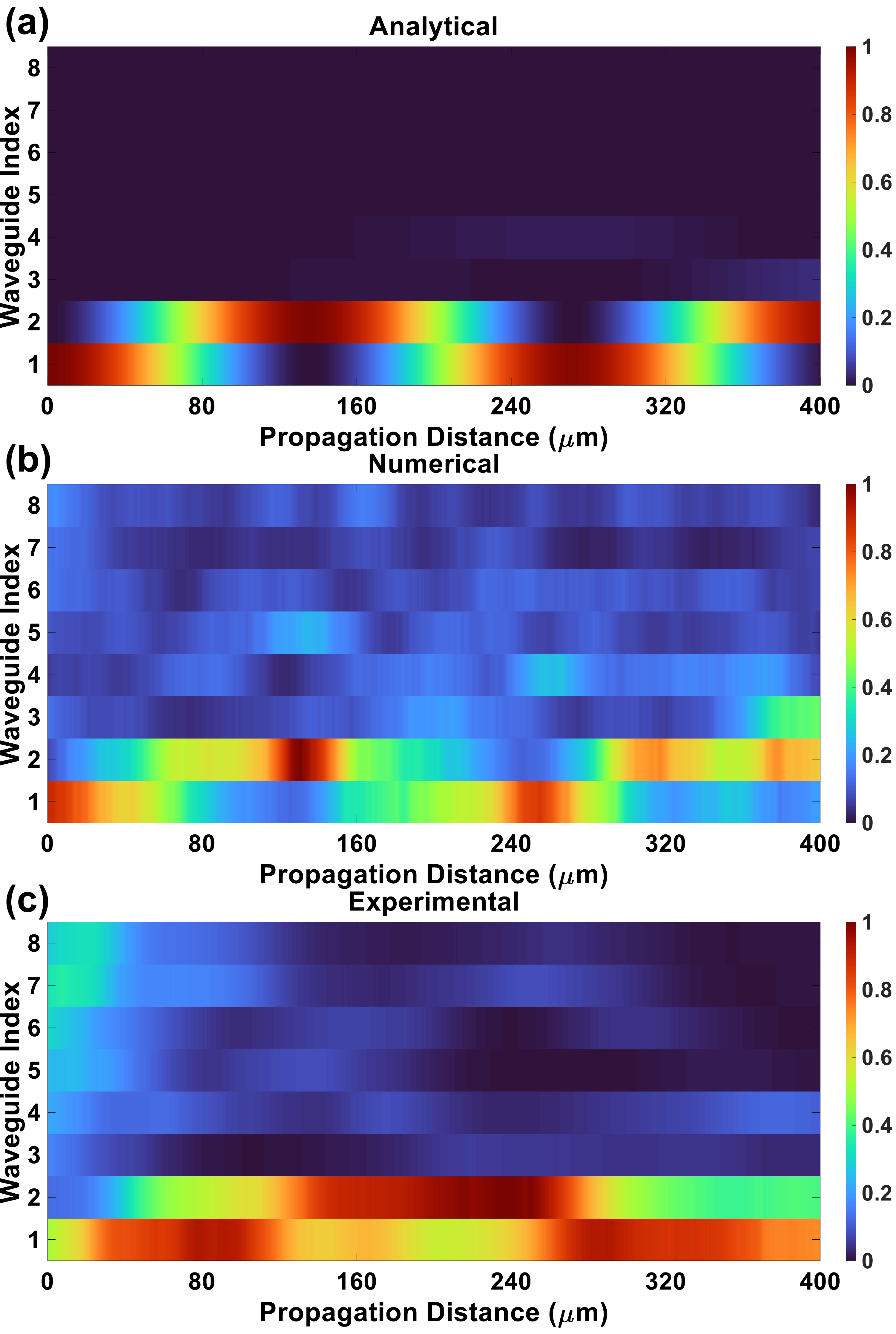}
\caption{Spatial intensity evolution map in the trivial ($\kappa_1 > \kappa_2$) lattice, consisting of eight coupled waveguides with excitation injected at the edge waveguide (Waveguide Index = 1), shown for (a) analytical, (b) numerical, and (c) experimental results.}
\label{fig:TrivialIntensities}
\end{figure}
It is established that no localization occurs in the absence of dimerization, but a bandgap is introduced through breaking the coupling symmetry in the coupled-waveguide systems, giving rise to topological and trivial lattice configurations. The system is now reduced to a system of eight coupled waveguides and coupling gaps of 2/12 $\mu$m are introduced to break the coupling symmetry ($\kappa_1 \neq \kappa_2$) with values taken from the relationship between $\kappa$ and $g$ in Fig.~\ref{fig:CouplingAnalysis}(f). Diagonalizing the system Hamiltonian yields the physical wavenumber in the propagation direction ($\beta_y$) associated with each of the waveguides and their eigenmodes in the system. These values are obtained from adding the single-waveguide wavenumber ($\beta_{y,o}$) to each eigenvalue of the Hamiltonian. Dimerization opens a bandgap in the wavenumber spectrum, but only in the topological configuration ($\kappa_1 < \kappa_2$) does this gap host two topologically protected edge states, arising as zero-energy solutions of the Hamiltonian that are fixed to the single ridge waveguide wavenumber, protected through chiral symmetry. The resulting spectrum is shown in Fig.~\ref{fig:eigenstates}(a), as well as the numerical mode profile for the edge states given the 2/12 $\mu$m gaps. The spatial signature is shown in Fig.~\ref{fig:eigenstates}(b), where the corresponding eigenvector amplitude at each waveguide index shows the strong localization that occurs in the edge waveguides only in the topological configuration, whereas bulk spreading occurs in the trivial case.

The same analytical and numerical framework is used to determine the mode spatial signature over the system's propagation length. Here, the coupling gaps are accordingly changed to introduce the contrast in $\kappa$ to introduce alternating strong and weak coupling between acoustic waveguides. Excitation is changed to the edge waveguide to visualize the expected edge state localization in the topological configuration, as opposed to the spreading into the bulk for the trivial. The analytical solution showing propagation dynamics for the edge-waveguide injection in the trivial configuration is shown in Fig.~\ref{fig:TrivialIntensities}(a). In the case of a trivial coupled waveguide system, the injected mode couples into the bulk of the structure and further spreads as the propagation length is increased. 

The structure is then constructed in COMSOL for numerical analysis in the same way it was for the uniform case. The numerically computed solution shows similar behavior, where there is spreading into the bulk. Again, a cut line is defined along the center of each waveguide in the direction of propagation, to show the spatial intensity map for the numerical solution in Fig.~\ref{fig:TrivialIntensities}(b). Spreading into the bulk is again observed, though to a greater extent than predicted analytically, likely due to effects not captured with idealized CMT. 

As with the bulk spreading case, the measured out-of-plane displacement is plotted as a spatial intensity map, shown in Fig.~\ref{fig:TrivialIntensities}(c), further showing spreading into the bulk. It should also be noted that a nonzero mode presence is observed at the input of every waveguide and not solely the one being injected. This is attributed to acoustic energy that is not fully injected into the input waveguide. This unwanted presence is spatially broad but decays rapidly with propagation distance as it is propagating away from the structure through the substrate. Together, these results confirm that dimerization does not guarantee localization, as the trivial configuration continues to spread into the bulk, in contrast to the topological configuration examined next.

\subsection{Topological Lattice (SSH)}
\begin{figure}[b]
\centering
\includegraphics[width=\linewidth]{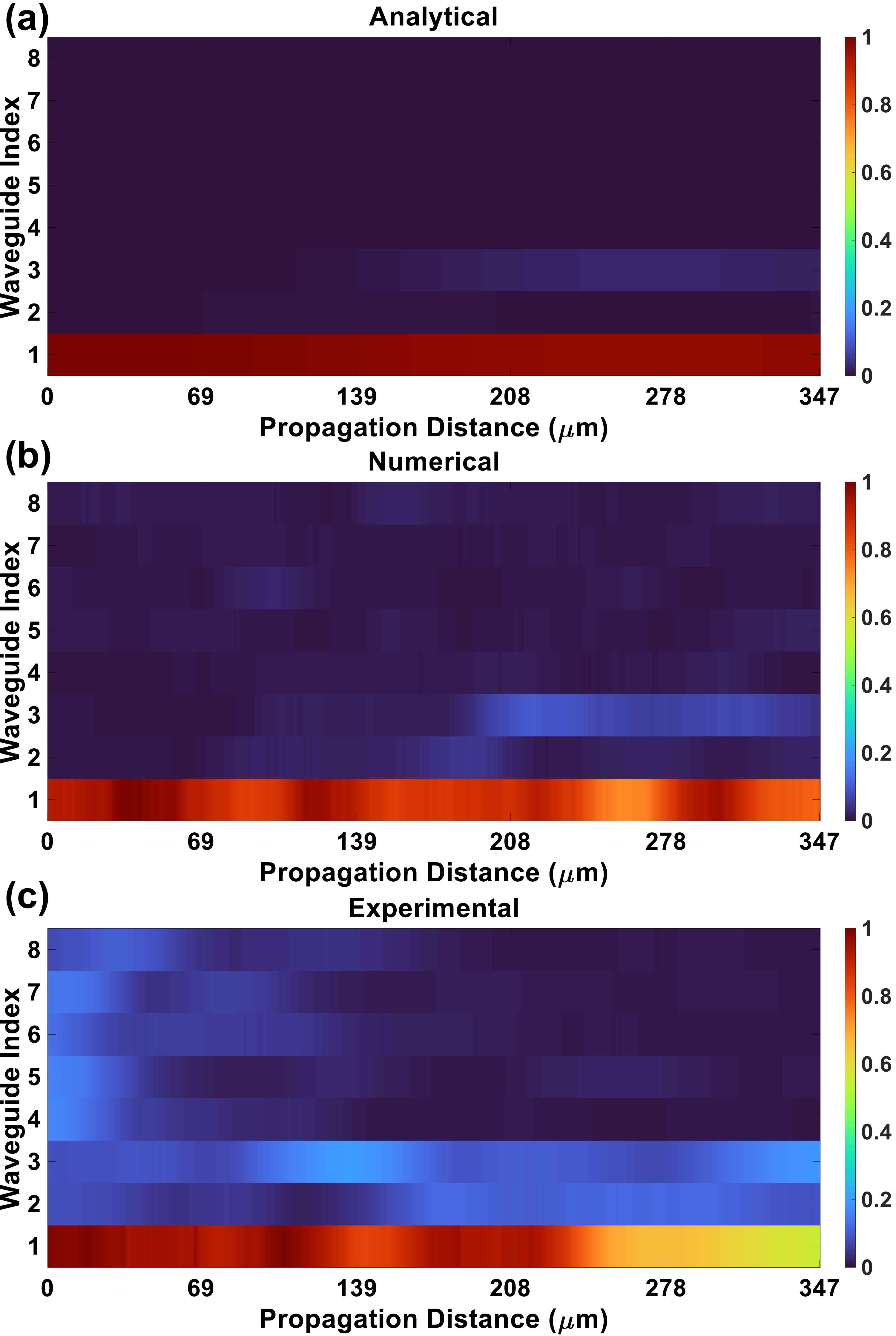}
\caption{Spatial intensity evolution map in the topological ($\kappa_1 < \kappa_2$) lattice, consisting of eight coupled waveguides with excitation injected at the edge waveguide (Waveguide Index = 1), shown for (a) analytical, (b) numerical, and (c) experimental results.}
\label{fig:TopologicalIntensities}
\end{figure}
As established previously, dimerization opens a bandgap, but only in the topological ($\kappa_1 < \kappa_2$) configuration does there exist protected edge states. The analysis now focuses on the spatial dynamics of the localization introduced by the SSH model. Excitation is applied to the edge waveguide using 12/2 $\mu$m coupling gaps that distinguish this configuration from the trivial counterpart. The analytical solution, shown in Fig.~\ref{fig:TopologicalIntensities}(a), shows strong localization of the injected mode to the edge waveguide, which is maintained across the full 350 $\mu$m propagation length. This shows a direct contrast to the bulk spreading seen in the trivial case. 

The structure is constructed in COMSOL for numerical analysis, following the same approach used for the trivial configuration. The numerically computed solution, shown in Fig.~\ref{fig:TopologicalIntensities}(b), shows extremely close agreement and replicates the strong edge state localization predicted analytically. This framework confirms that topological protection is captured in full wave simulation. 

Experimentally, the measured out-of-plane displacement for the topological configuration is visualized as a spatial intensity map in Fig.~\ref{fig:TopologicalIntensities}(c). These results confirm strong and sustained localization to the edge waveguide, while staying in close agreement with analytical and numerical predictions. As with the trivial case, a nonzero mode presence is observed at the input of every waveguide due to acoustic energy that is not fully injected into the input waveguide. However, this presence decays rapidly with propagation distance. Together, these results demonstrate that the topological dimerization produces a topological edge state localization that is robust and consistent across analytical, numerical, and experimental methods.

\subsection{Thouless Pumping}
As previously seen, the static SSH topological configuration pins a localized edge state to an engineered boundary, but the state itself does not move within the system of coupled waveguides. Thouless pumping is used to conserve topological properties while also transporting the localized mode across the system, from one topologically protected edge state to another \cite{citro_thouless_2023}. This is achieved by varying the coupling coefficient along the propagation direction, made parametric with a synthetic pumping phase, $\phi(y)$. The system Hamiltonian follows that of the Aubry-Andr\'e-Harper model, denoted by Eq.~\ref{eq:H_phi}.

\begin{equation}
\label{eq:H_phi}
    \mathbf{H}(\phi) = \sum_{m=1}^{N-1} \kappa_{m,m+1}(\phi)\left(|m\rangle\langle m+1| + \text{h.c.}\right)
\end{equation}

Here, the coupling coefficient $\kappa(\phi)$ is described by Eq.~\ref{eq:kappa_phi}, where $\kappa_o$, $\lambda$, $b$, and $\phi$ represent the average coupling coefficient throughout the system, the modulation depth, modulation periodicity, and pumping phase, respectively. Physically, $\lambda$ controls how strongly the coupling coefficient deviates from $\kappa_o$ between waveguides. Additionally, the periodicity parameter, $b$, sets how many waveguides it takes for the coupling pattern to repeat itself.

\begin{equation}
\label{eq:kappa_phi}
    \kappa_{m,m+1}(\phi) = \kappa_0\left(1 + \lambda\cos(2\pi b m + \phi)\right)
\end{equation}

In order to keep the system as simple and compact as possible, it was chosen to have $N=10$ waveguides, $b=2/3$, and an average coupling gap (and corresponding coupling coefficient, $\kappa_o$) of 4 $\mu$m. The modulation depth was set to $\lambda=0.7$, the largest value consistent with fabrication constraints (yielding a minimum gap size of 1.94 $\mu$m), to maximize $\kappa$ contrast across the pump cycle. The choice of $b$ is constrained by the requirement that the bulk gap remain open throughout the pump cycle. Choosing $b=1/2$ reduces the coupling modulation to the same two-site periodicity as the SSH implementation, forcing the bulk gap to close at $\phi=\pi/2$, while instead choosing $b=2/3$ allows a gap to remain open across the entire pump cycle, avoiding this forced closure and spreading into the bulk, further described in Section III of the Supplementary Material.

\begin{figure}[b]
\centering
\includegraphics[width=\linewidth]{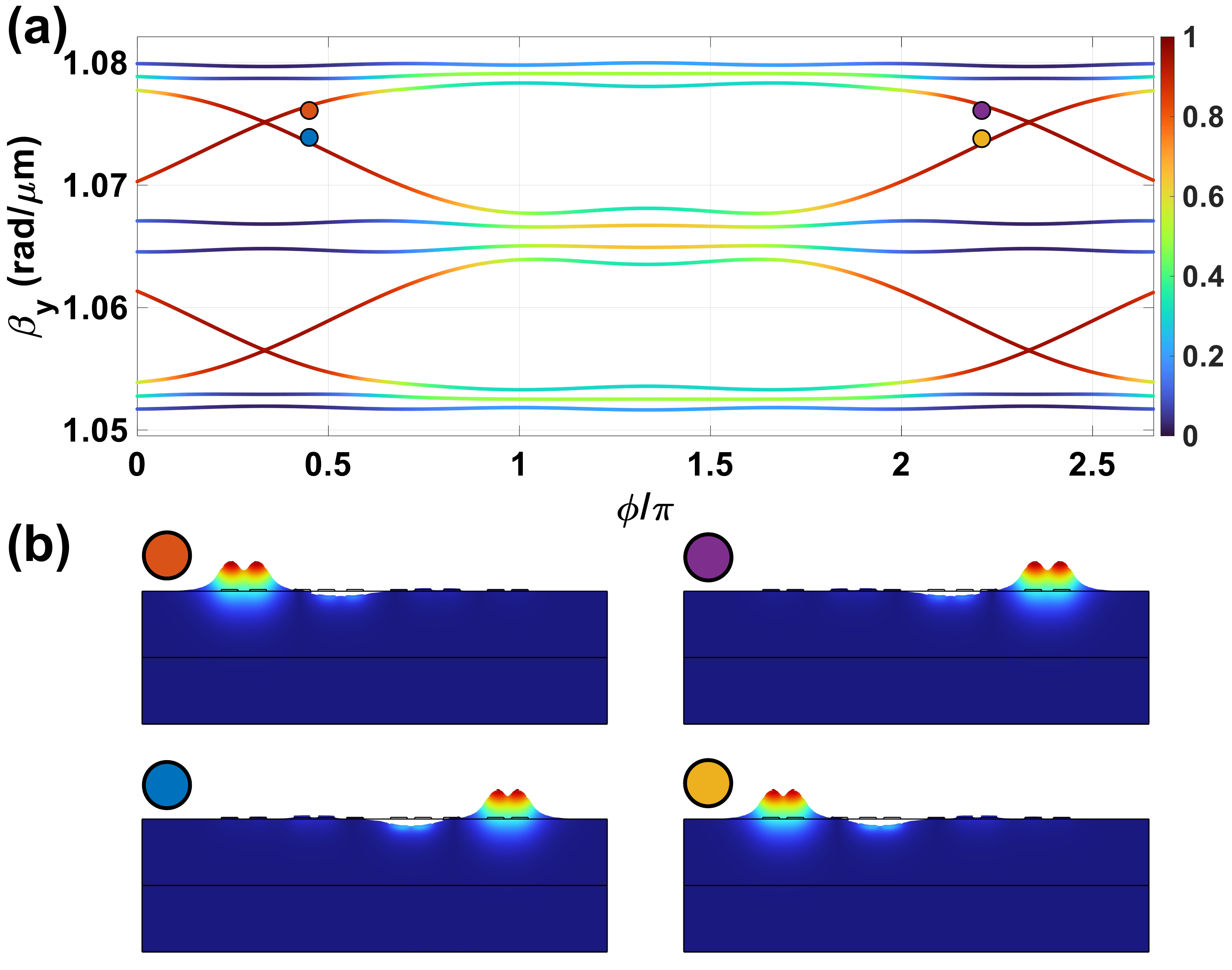}
\caption{(a) Dispersion highlighting $\beta_y$ over pumping parameter ($\phi$) for Thouless pumping in a 10 coupled-waveguide system, with color indicating localization weight (0 = bulk, 1 = fully edge localized). (b) The colored markers indicate the mode shapes indicated at the desired $\phi_{start}$ = 0.45$\pi$ and $\phi_{stop}$= 2.21$\pi$, highlighting the shift in edge state from one to the opposite as $\phi$ is varied.}
\label{fig:PumpingBands}
\end{figure}

The pump trajectory is determined by sweeping the pumping phase and diagonalizing the Hamiltonian to obtain its eigenvalues ($\lambda_n(\phi)$) and eigenvectors ($\vec{\nu}_n(\phi)$) for each eigenstate index, $n$, where $\nu_{n,m}(\phi)$ denotes the $m^{th}$ component of $\vec{\nu}_n(\phi)$. In other words, this represents the amplitude of eigenstate $n$ at waveguide $m$, giving an indication of mode presence in each waveguide. Additionally, each wavenumber is computed by $\beta_{n,y}(\phi) = \beta_{y,o} + \lambda_n(\phi)$.  For each eigenvector, a weight is assigned to measure the localization to the outermost two waveguides, given by Eq.~\ref{eq:weights}.

\begin{equation}
\label{eq:weights}
w_n(\phi) = \sum_{m=1}^{2}\left|\nu_{n,m}(\phi)\right|^2 + \sum_{m=9}^{10}\left|\nu_{n,m}(\phi)\right|^2
\end{equation}

Since the pumping phase is a synthetic parameter, it is swept analytically. From this swept parameter ($\phi$) and the system's other defined parameters, the band diagram is constructed, with color indicating the edge localization shown in Fig.~\ref{fig:PumpingBands}(a). The solid markers represent specific $\phi$ values and their corresponding COMSOL mode shapes at $\phi$ = 0.45$\pi$ and 2.21$\pi$, corresponding to strongly localized edge states, with the respective mode shapes shown in Fig.~\ref{fig:PumpingBands}(b). As $\phi$ is swept, the structure transitions from an edge state with strong localization, through the bulk, and arrives at an edge state on the other boundary of the system. Ultimately, the device is designed such that $\phi$ varies adiabatically along the propagation direction from $\phi_{start}$ = 0.45$\pi$ to $\phi_{stop}$ = 2.21$\pi$, transporting the injected mode across the structure to the opposite edge with strong localization.

\begin{figure}[t]
\centering
\includegraphics[width=\linewidth]{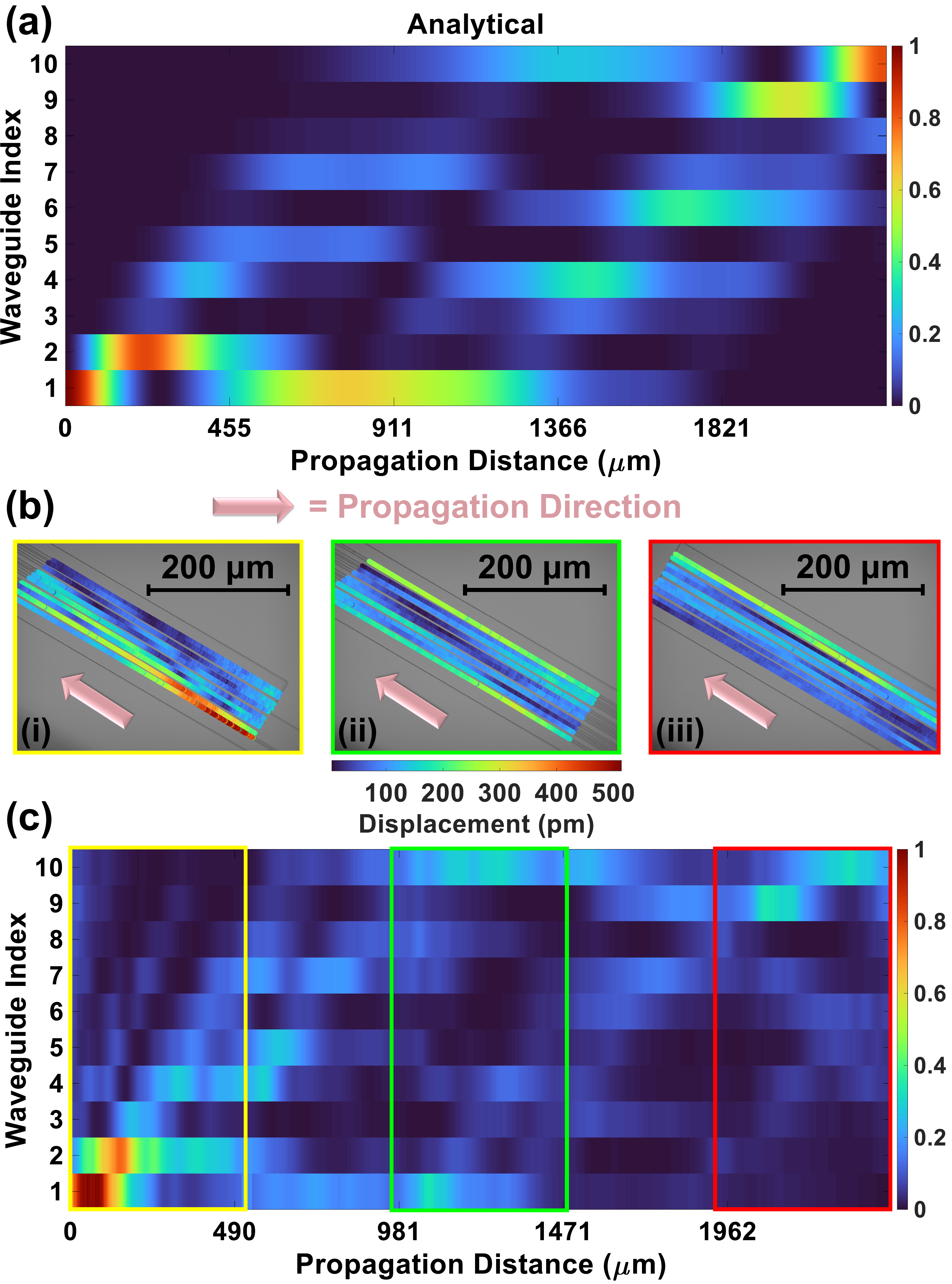}
\caption{(a) Spatial intensity evolution for the Thouless pumping structure, consisting of 10 coupled waveguides with excitation injected at the edge waveguide (Waveguide Index = 1), shown for analytical results. (b) Fabricated pumping structure overlaid with individual LDV scans at the (i) beginning, (ii) middle, and (iii) end of the device, showing the measured out-of-plane displacement. The five scans required to image the full device are stitched together in (c), where the resulting spatial intensity map shows the transport of the edge state from the input waveguide to the opposite boundary.}
\label{fig:PumpingIntensities}
\end{figure}

The values of the synthetic parameter must be mapped to physical gaps within in the system. Here, the Hamiltonian  is reconstructed according to Eq.~\ref{eq:H_phi} and Eq.~\ref{eq:kappa_phi} at each waveguide index given for a given $\phi$, yielding the specific $k_{m,m+1}$. This is then inverted to find its coupling gap for that specific $\phi$, ultimately mapping the pumping phase to a physical gap. Gradual tapering is introduced between neighboring gaps corresponding to necessary $\kappa$ to map $\phi$ to propagation length into a slowly varying physical taper. The length of the total structure was chosen to be 10$L_c$, where $L_c$ is the coupling length for the system-defined average coupling coefficient, $\kappa_o$, ultimately pumping the injected mode at one edge state to its opposite boundary edge state. The analytical formulation is shown in Fig.~\ref{fig:PumpingIntensities}(a), highlighting the spatial intensity evolution and transport of the mode from one edge state to the opposite boundary. A full numerical simulation of this process is not computationally feasible as the tapering length is too large to simulate directly. 

Having established that alternating weak/strong gaps are sufficient to localize a SAW to the edge of a static topological lattice, the SSH model was successfully implemented in the acoustic domain. This is now extended in the ScAlN on SiC platform to demonstrate Thouless pumping, in which an injected fundamental Rayleigh mode is dynamically transported from the edge state at one boundary to the edge state at the opposite boundary. This pumping utilizes the topological invariance of the system to implement robust and quantized transport of the injected edge mode across the system lattice.  Building on the theoretical formulation, the pumping trajectory is experimentally realized by varying the coupling coefficients between neighboring waveguides by varying the pumping parameter from a designed $\phi_{start}$ = 0.45$\pi$ to $\phi_{stop}$ = 2.21$\pi$, characterized through LDV measurement.

Since the device was fabricated to have a total length of 10$L_c$, the pumping structure exceeded the field of view for a single LDV scan and required five overlapping scans to be stitched together to capture the full pumping process. Individual scans are shown in Fig.~\ref{fig:PumpingIntensities}(b) at the (i) beginning, (ii) middle, and (iii) end of the pumping structure, with the measured out-of-plane displacement overlaid on the fabricated device. At the input, the displacement is strongly localized to the bottom-most boundary of the system when the fundamental Rayleigh mode is injected into the edge state with pumping phase $\phi$ = 0.45$\pi$. Moving through the central waveguides, the mode intensity becomes distributed amongst the central waveguides as the coupling gaps are adiabatically tapered and moving the system through varying pumping phase. At the output of the device, the displacement was maximum at the upper boundary at the system, with no presence of the mode at the lower, indicating a successful transport to $\phi$ = 2.21$\pi$.

The five scans are stitched together into a continuous spatial intensity map, shown in Fig.~\ref{fig:PumpingIntensities}(c). This map shows the pumping trajectory and the transport from the injected waveguide to the waveguide at the opposite boundary, in excellent agreement with the predicted analytical solution. This measurement represents direct experimental validation that Thouless pumping of a SAW mode can be implemented in acoustic waveguides, completing the theoretical, numerical, and experimental analysis established.

\begin{table*}[!ht]
\centering
\caption{Solution similarity across analytical (A), numerical (N) and experimental (E) results, quantified through inner products ($\chi_{A,N}$, $\chi_{A,E}$, and $\chi_{N,E}$).}
\label{tab:overlap}
\renewcommand{\arraystretch}{1.3}
\setlength{\tabcolsep}{12pt}
\begin{tabular}{lccc}
\toprule
\textbf{Configuration} & \textbf{Analytical/Numerical} & \textbf{Analytical/Experimental} & \textbf{Numerical/Experimental} \\
\midrule
Topological & 0.993    & 0.934 & 0.952 \\
Trivial     & 0.743    & 0.616 & 0.637 \\
Uniform     & 0.683    & 0.629 & 0.807 \\
Pumping     & $\times$ & 0.636 & $\times$ \\
\bottomrule
\end{tabular}
\end{table*}

\section{Similarity Analysis}

To quantify the agreement between the analytical, numerical, and experimental results, a solution overlap, $\chi$ is defined and calculated between each respective case. Given by Eq.~\ref{eq:similarity}, the metric takes values between 0 and 1, consisting of the inner product between two normalized mode profiles $\psi_i$ and $\psi_j$. A value of $\chi$ = 1 indicates perfect accuracy between solutions and $\chi$ = 0 indicates no spatial overlap between solutions.

\begin{equation}
\begin{split}
\chi_{i,j} &= \left| \langle \psi_i | \psi_j \rangle \right|^2 \\
&= \left| \frac{\int \psi_i^*(y)\, \psi_j(y)\, dy}{\sqrt{\int |\psi_i(y)|^2\, dy} \sqrt{\int |\psi_j(y)|^2\, dy}} \right|^2
\end{split}
\label{eq:similarity}
\end{equation}

In Eq.~\ref{eq:similarity}, $\psi_i$ and $\psi_j$ correspond to the mode profiles obtained from the analytical, numerical, and experimental methods. The results are summarized in Table ~\ref{tab:overlap}. The topological configuration shows the strongest agreement across all three methods with similarity values of $\chi_{A,N}$ = 0.993, $\chi_{A,E}$ = 0.934, and $\chi_{N,E}$ = 0.952. This indicates that the topologically protected edge state is theoretically robust and captured across full-wave numerical simulation and physical measurements, proving that the acoustic wave edge state localization is a measurable feature of phononic devices. Alternatively, the trivial ($\chi_{A,N}$ = 0.743, $\chi_{A,E}$ = 0.616, and $\chi_{N,E}$ = 0.637) and uniform spreading case ($\chi_{A,N}$ = 0.683, $\chi_{A,E}$ = 0.629, and $\chi_{N,E}$ = 0.807) show lower agreement across all methods. This is a physically meaningful trend, indicating the spreading into the bulk with more real-world artifacts like fabrication and material defects or mode injection initial conditions, a feature that the topological case is robust against. This indicates that the topological localization is more reproducible due to the protection against perturbations. The Thouless pumping configuration, limited to $\chi_{A,E}$ shows a similarity of 0.636, on the order of the other solutions. Additional inaccuracy is largely attributed to the scan area needed to capture the full behavior of the pumping device, in which the solutions needed to be manually stitched together, and the larger propagation length of the device in which propagation loss significantly contributes. Contributions due to propagation loss ($\alpha$) lead to a modified version of Eq.~\ref{eq:analyticalSolution}, $\vec{a}\left(y\right) = e^{-j(\beta_y-j\alpha) y}e^{-j\mathbf{H}y}\vec{a}\left(0\right)$. Including a propagation loss of 1.6 dB/mm \cite{guida_phononic_2025}, the solution accuracy improves to $\chi_{A,E}$ = 0.703. Still, the solution similarity indicates that mode transport behavior between edge state is occurring in agreement with theoretical behavior. 

Together, these results demonstrate that theoretical, numerical, and experimental methods not only validate each other, but do so with convincing qualitative accuracy that improves with and emphasizes topological protection. 

\section{\label{sec:Conclusion} Conclusion}
Overall, this work establishes a working analytical, numerical, and experimental framework for implementing topological principles in traveling SAW acoustic waveguide based devices in the ScAlN on SiC platform. Uniform coupling lattice structures were first implemented to show the behavior of the injected fundamental Rayleigh mode spreading into the bulk with the absence of dimerization. Dimerization was then introduced to realize the SSH model, producing topologically protected edge states with strong localization. Building on the static validation, Thouless pumping was demonstrated by adiabatically varying coupling gaps to achieve the transport of an injected topologically protected edge mode between opposite boundaries of the lattice. For the first time, this work systematically addresses the validity of experimentally measured results against that of theoretical and numerical approaches across all configurations, showing excellent accuracy between the different methods. Here, accuracy between approaches exceeded 90\% for the topologically protected configurations, reaching as high as 99\%. This concept can be adapted to other slow-on-fast piezoelectric platforms that are compatible with optical mode integration, such as the ScAlN on sapphire platform \cite{guida_low-mathrmv_pi_2026}. Future work will extend this framework toward topologically protected optomechanical waveguides, leveraging the demonstrated robustness of acoustic edge states and pumping techniques to enable protected coupling between acoustic and optical domains for applications in non-reciprocal signal processing and topologically protected optomechanical transduction.

\section{\label{sec:Ackowledgement} Acknowledgment}
This material is based upon work supported by the Air Force Office of Scientific Research under award number FA9550-23-F-0014, the DARPA Young Faculty Award (YFA) under Grant D24AP00005, and the NSF CAREER under Award 2340405.
\section{\label{sec:DataAvailability} Data Availability}
All data generated or analyzed during this study are included in this published article and its supplementary material.
\section{Author Contributions}
J.G., S.M., and S.G. conceptualized the study. J.G. fabricated the devices, developed the analysis code, and performed all analytical, numerical, and experimental analysis. N.P. performed the LDV measurements. J.G. wrote the manuscript, with N.P. contributing to all sections. S.G. supervised the project. S.G., S.M. and C.C. provided feedback on the manuscript and contributed to revisions.
\section{Competing Interests}
The authors declare no competing interests
\section{Correspondence}
Correspondence and requests for materials should be addressed to
Siddhartha Ghosh.


\bibliography{Bibliography}

\end{document}